\documentclass[aps,prl,twocolumn,showpacs,showkeys,footinbib,superscriptaddress,longbibliography]{revtex4-1}

\usepackage{amsmath}
\usepackage{amssymb}
\usepackage{graphicx}
\usepackage{bm}
\usepackage{color}
\usepackage{relsize}
\usepackage{braket}
\usepackage[caption=false]{subfig}

\usepackage{hyperref}
\usepackage[all]{hypcap}

\begin{document}
\title{Rashba spin-orbit coupling enhanced magnetoresistance in junctions with one ferromagnet}
\author{Chenghao Shen}
\affiliation{Department of Physics, University at Buffalo, State University of New York, Buffalo, NY 14260, USA} 
\author{Ranran Cai}
\affiliation{International Center for Quantum Materials, School of Physics, Peking University, Beijing 100871, P. R. China}
\author{Alex Matos-Abiague}
\affiliation{Department of Physics and Astronomy, Wayne State University, Detroit, MI 48201, USA}
\author{Wei Han}
\affiliation{International Center for Quantum Materials, School of Physics, Peking University, Beijing 100871, P. R. China}
\author{Jong E. Han}
\affiliation{Department of Physics, University at Buffalo, State University of New York, Buffalo, NY 14260, USA}
\author{Igor \v{Z}uti\'c}
\affiliation{Department of Physics, University at Buffalo, State University of New York, Buffalo, NY 14260, USA}
\date{\today}
\begin{abstract}
We explain how Rashba spin-orbit coupling (SOC) in a two-dimensional electron gas (2DEG), 
or in a conventional $s$-wave superconductor, can lead to a large magnetoresistance even with one ferromagnet. However, such enhanced magnetoresistance 
is not generic and can be nonmonotonic and change its sign with Rashba SOC. For an in-plane rotation of magnetization, it is typically negligibly small for a 2DEG and depends on the perfect transmission which emerges 
from a spin-parity-time symmetry of the scattering states, while this symmetry is generally absent from the Hamiltonian of the system. 
The key difference from considering the normal-state magnetoresistance is the presence of the spin-dependent Andreev reflection at superconducting interfaces.
In the fabricated junctions of quasi-2D van der Waals ferromagnets with conventional $s$-wave superconductors (Fe$_{0.29}$TaS$_2$/NbN) we find another 
example of enhanced magnetoresistance where the presence of Rashba SOC reduces the effective interfacial strength and is responsible for an equal-spin Andreev reflection. 
The observed nonmonotonic trend in the out-of-plane magnetoresistance with the interfacial barrier is an evidence for the proximity-induced equal-spin-triplet superconductivity. 
\end{abstract}

\maketitle

\section{I. Introduction}
The magnetoresistance (MR) is a key figure of merit in spintronics and its enhancement is associated with the major advances
in magnetically sensing and storing information~\cite{Maekawa1982:ITM,Moodera1995:PRL,Zutic2004:RMP,Parkin2004:NM,Tsymbal:2019}. 
Typically, a large MR is sought in structures with 
multiple ferromagnetic regions, where the resulting spin-valve effect implies that the resistance of the systems depends of the relative magnetization, {\bf M}
orientation of those ferromagnets. 

However, as first discovered in 1857 by Lord Kelvin~\cite{Thomson1857:PRSL}, anisotropic MR (AMR)
shows that with spin-orbit coupling (SOC) there is a change of the electrical resistivity with the relative direction of the charge current 
with respect to {\bf M} of a single bulk ferromagnet (F), such as Ni or Fe. Another example of MR with a single F region is the tunneling AMR (TAMR)~\cite{Gould2004:PRL,Moser2007:PRL,Chantis2007:PRL,Fabian2007:APS}, also a manifestation of 
the interplay between SOC and {\bf M}. Unfortunately, while a single F region simplify scaled-down devices 
both AMR and TAMR are limited by their small magnitudes (typically $<1\,\%$ for in-plane {\bf M} rotation)~\cite{Shen2020:PRB,Fabian2007:APS}.

\begin{figure}[t]
\vspace{-1cm}
\centering
\includegraphics*[width=9.3cm]{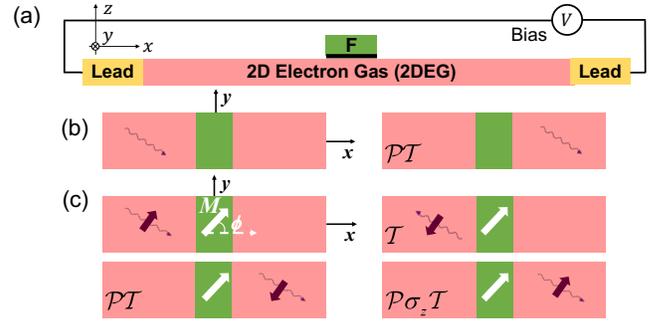}
\vspace{-2.1cm}
\caption{(a) Lateral geometry,  
the current flows in the 2DEG. 
(b) Action of the time reversal, $\mathcal{T}$,  and  the space inversion operator, $\mathcal{P}$,  on the incident wave. 
By applying the $\mathcal{P} \mathcal{T}$ operator, the incident wave on the left is transformed into a transmitted wave on the right. For a spinless barrier system,
the $\mathcal{P} \mathcal{T}$  symmetry gives  the perfect transmission. 
(c) Action of the $\mathcal{P}\sigma_z\mathcal{T}$ 
operator and the $\mathcal{P}\sigma_z\mathcal{T}$ symmetry  generalizes the condition from (b) for an incident wave with an in-plane spin and 
a barrier with the in-plane  magnetization ${\mathbf M}$, defined by the polar angle $\phi$.}
\label{fig:setup}
\end{figure}

In this work we explore a possibility for a much larger MR with a single F region and Bychkov-Rashba, also known as Rashba SOC~\cite{Bychkov1984:JETPL},
in both normal and superconducting state. The resulting enhanced MR is not only significant for the potential spintronic 
applications, but could also distinguish between the trivial and topological states~\cite{Shen2020:PRB}, or provide a signature of equal-spin-triplet 
superconductivity, sought to realize coexistence of ferromagnetism and superconductivity, dissipationless spin currents, and
Majorana bound states for fault-tolerant topological quantum 
computing~\cite{Cai2022:AQT,Amudsen2022:P,Eschrig2015:RPP,Linder2015:NP,Buzdin2005:RMP,Bergeret2005:RMP,Kitaev2001:PU,%
Aasen2016:PRX,Gungordu2022:JAP,Laubscher2021:JAP}.  
 
To realize an enhanced normal-state MR, we consider 
a lateral geometry from Fig.~\ref{fig:setup}(a). The resulting TAMR is not
determined by highly spin-polarized carriers or a large exchange energy in the F region, but by the 
high interfacial transmission
from the interplay of the Rashba SOC, the barrier strength at the F/2DEG interface, and
the proximity-induced exchange field, which can reach tens of meV 
and extend over tens of nm~\cite{Zutic2019:MT,Akimov2017:PRB,Korenev2019:NC,Takiguchi2019:NP,Betthausen2012:S}.
 
For a simple spinless case of a square barrier with thickness $d$ and height $V_0$, a standard expression for transmission, $T$,  with energy, $E > V_0$, is~\cite{Griffiths:2005,Sakurai:2011}
\begin{equation}
T=\left(1+V_0^2 \sin^2(kd)/[4E(E-V_0)] \right)^{-1},
\label{eq:T}
\end{equation}
where $k=\sqrt{2m(E-V_0)}/\hbar$
is the wave vector. The perfect transmission, $T=1$, requires
(i) $V_0 = 0$
or (ii) $kd=n\pi$, where $n=1,2,3,...$, gives the resonance condition. This well-known behavior~\cite{Griffiths:2005,Sakurai:2011}
is usually not connected to the parity-time symmetry $\mathcal{P}\mathcal{T}$, depicted in Fig.~\ref{fig:setup}(b), which 
satisfies both cases (i) and (ii). Such a symmetry,
where $\mathcal{P}$ is the parity operator,
$\mathcal{T} = \mathcal{K}$ is the time-reversal operator, 
and $\mathcal{K}$ is the complex conjugation operator, which ensures that the incoming and transmitted spinless wave are identical, up to a phase. 

However, generalization of  the perfect transmission for spin-1/2 carriers, where 
$\mathcal{T} =  - i{\sigma _y}\mathcal{K}$, and $\sigma _y$ is the Pauli matrix, is much less explored
with SOC and magnetic barriers.
It was recently found that the perfect transmission emerges when the eigenstates of the 
F/2DEG Hamiltonian, $H_\text{F/2DEG}$, satisfy 
$\mathcal{P}\sigma_z\mathcal{T}$, the spin-parity-time symmetry, where 
[$\mathcal{P}{\sigma _z}\mathcal{T}$, $H_\text{F/2DEG}$]$\neq0$~\cite{Shen2020:PRB}.  
Intuitively, the emerging perfect transmission for the eigenstates of  $H_\text{F/2DEG}$ 
which satisfy $\mathcal{P}\sigma_z\mathcal{T}$ 
can be understood from Fig.~\ref{fig:setup}(c).
$\mathcal{T}$ reverses the spin and motion of the incident wave, while $\mathcal{P}\sigma_z$ 
inverts both the spin (through $\sigma_z$) and position (through $\mathcal{P}$) of the wave. 
By applying the $\mathcal{P}\sigma_z\mathcal{T}$ operator the incident wave on the left is transformed to itself, but as a transmitted wave on the right. 
Therefore, scattering states which are eigenfunctions of $\mathcal{P}\sigma_z\mathcal{T}$ experience perfect transmission. 
The resulting in-plane TAMR amplitude 
\begin{equation}
{\rm TAMR_\|}=\frac{G(\phi=0)-G(\phi=\pi/2)}{G(\phi=\pi/2)},
\label{eq:tamrIN}
\end{equation}
where the $\phi$-dependent conductance $G$ is the inverse of the resistance, $G=1/R$, with $\phi$ defined in Fig.~1(c),
can be enhanced by one or two orders of magnitude.

To study an enhanced out-of-plane MR in the superconducting state, we focus on the recent transport experiments in 
junctions of quasi-2D van der Waals F with conventional $s$-wave superconductors (S) (Fe$_{0.29}$TaS$_2$/NbN)~\cite{Cai2021:NC}.
Compared to the normal state MR, the key difference is the presence of Andreev reflection at the F/S interface with Rashba SOC.
In this process an incoming electron is reflected as a hole and the Cooper pair enters the S region. With 
SOC, in addition
to the conventional Andreev reflection, where the incoming and reflected particle have an opposite spin, a spin-flip or equal-spin Andreev reflection 
is also possible in which the incoming and reflected particles have the same spin~\cite{Zutic1999:PRBa,Hogl2015:PRL}. While the first process 
is responsible for the spin-singlet proximity-induced superconductivity, the second yields the spin-triplet counterpart. 

\begin{figure}[t]
\vspace{-0.7cm}
\centering
\includegraphics*[trim=0cm 0.4cm 0.2cm 1.4cm,clip,width=8.7cm]{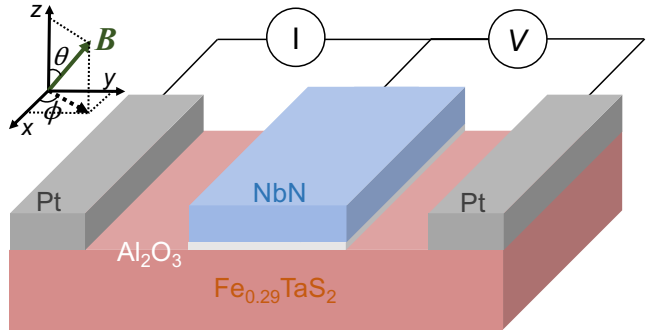}
\vspace{-1.1cm}
\caption{Schematic of the geometry including a quasi-2D vdW F/S junction Fe$_{0.29}$TaS$_2$/NbN, 
with an Al$_2$O$_3$ insulating barrier. MR measurements are performed for an out-of-plane ${\mathbf M}$ rotation, 
defined by angle $\theta$ with respect to the applied magnetic field, {\bf B}. The interfacial resistance is obtained as 
$V/I$.}
\label{fig:FS}
\end{figure}

The corresponding anisotropic behavior, magnetoanisotropic Andreev reflection (MAAR) can be viewed  as the generalization of 
TAMR~\cite{Hogl2015:PRL,Vezin2020:PRB,Lv2018:PRB}, which is recovered in the normal state of the F/S junction for a large bias, $V$, 
applied magnetic field, $B$, or a high temperature. 
From a full MAAR calculation for the F/S junction, such a TAMR value can be obtained by taking the 
superconducting gap to vanish. 
Similar to Eq.~(\ref{eq:tamrIN}), the out-of-plane MAAR amplitude in F/S junctions is 
\begin{equation}
{\rm MAAR_\bot}=\frac{G(\theta=0)-G(\theta=\pi/2)}{G(\theta=\pi/2)},
\label{eq:tamrOUT}
\end{equation}
where $\theta$ is the angle defined in Fig.~\ref{fig:FS}. As we later show, F/S measurements for the MAAR can be orders
of magnitude larger than TAMR in the normal state. This enhancement is dominated by the spin-flip Andreev reflection with
Rashba SOC at the F/S interface. 

Despite the different geometries for the considered F/2DEG and F/S structures, which also differ by the in-plane vs out-of-plane rotation
of {\bf M}, we find important common features and surprising nonmonotonic trends in TAMR in both systems. Some of these identified 
trends arise from the role of Rashba SOC which modifies the effective barrier strength and determines the condition for perfect transmission
or an enhanced contribution of the spin-flip Andreev reflection. 

Following this Introduction, we explore the structures from Fig.~1(a). In Sec.~II we describe the F/2DEG Hamiltonian, consider the influence 
of SOC on the scattering states and analyze the conductance which reveals different resonant-like behavior, peaked at different parameters. The calculation of in-plane TAMR in Sec.~III reveals 
different trends as a function of the barrier and SOC strengths as well as the proximity-induced exchange splitting. 
In Sec.~IV 
we examine the superconducting structures from Fig.~2 and describe the corresponding F/S Hamiltonian, followed by 
the measured and calculated conductance using a simple F/S Hamiltonian. In Sec.~V we describe how the predicted influence of the Rashba
SOC on the effective barrier strength provides a good description of the measured out-of-plane MR and confirms the evidence
for the equal-spin-triplet superconductivity
 in the considered heterostructures. In Sec.~VI we provide additional discussion for the relevance of enhanced MR in the
 normal and superconducting state and note some open questions for future work.

\section{II. F/2DEG HAMILTONIAN AND CONDUCTANCE ANALYSIS}

\begin{figure}[t]
\vspace{-1.2cm}
\centering
\includegraphics*[trim=0cm 0cm 0cm 0cm,clip,width=11.2cm]{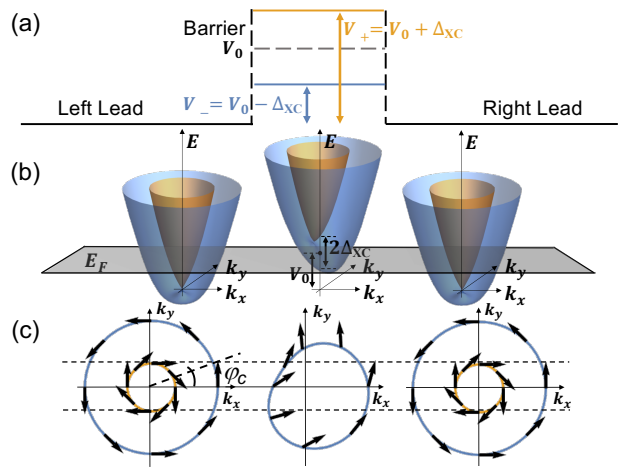}
\vspace{-1.8cm}
\caption{(a) Schematic of the spin-dependent barrier. Its effective strength is modified by the proximity-induced exchange
field $\Delta_\text{xc}$, shown in a weak SOC limit.
(b) 2DEG band structure with the Fermi energy $E_F$ and the effective barrier region (middle) of height $V_0$ and 
$\Delta_\text{xc}$. (c) Corresponding Fermi contours, arrows denote the spin orientation. Dashed lines: the range of a 
conserved wave vector $k_y$ in the scattering states. For incident angles exceeding $\varphi_c$ backscattering is suppressed.}
\label{fig:bands}
\end{figure}

To explore the interplay of the proximity-modified 2DEG and SOC, together with the magnetic anisotropy of the transport
properties, we consider a model Hamiltonian of the system represented in Fig.~\ref{fig:setup}(a) given by  
\begin{equation}
H_\text{F/2DEG} = \frac{p^2}{2m^\ast}+ \frac{\alpha}{\hbar}\left(\boldsymbol{\sigma}\times\mathbf{p}\right)\cdot\mathbf{\hat{z}}
+ [V_0 - \Delta_\text{xc}(\mathbf{m} \cdot \boldsymbol{\sigma})]h(x),
\label{eq:H2DEG}
\end{equation}
where  $m^\ast$ is the effective mass, $\alpha$ is the Rashba SOC strength, $\mathbf{\hat{z}}$ is the unit vector along the $z$-axis, $\mathbf{p}=(p_x,p_y)$ is the 2D momentum operator, $\boldsymbol{\sigma}$ is the vector of Pauli matrices, $V_0$ describes the potential barrier, $\Delta_\text{xc}$  and $\mathbf{m}$ are the magnitude and direction of the proximity-induced ferromagnetic exchange field. 
The function $h(x)=\Theta(d/2+x)\Theta(d/2-x)$ describes a square barrier of thickness $d$. 
We focus on electrons, not holes~\cite{Winkler:2003,Rozhansky2016:PRB,Liu2018:PRL,Arovas1998:PRB}, with the 
effective barrier region and typical band structure shown in Fig.~\ref{fig:bands}. In Fig.~\ref{fig:bands}(a) the resulting
spin-dependent barrier is shown in a weak SOC limit,   $\alpha k_F \ll \Delta_\text{xc}$, where $k_F = \sqrt{2 m^\ast E_F / \hbar}$ 
is the Fermi wave vector
averaged over the inner and outer contours in the leads, while  $E_F$ is the Fermi energy.
The blue (yellow) contours in Fig.~\ref{fig:bands}(c) denote lower (upper) bands.  Inside the barrier the 
upper band is irrelevant since its bottom is above $E_F$. The spin orientations are marked by arrows.

Due to  Rashba SOC, the wave functions can be classified by the helicity index, 
where $\lambda=1~(-1)$ refers to the inner (outer) Fermi contour as depicted in Fig.~\ref{fig:bands}(c).
The scattering states for the finite square barrier model can be written as ${\psi _\lambda }\left( {x,y} \right) = (1/\sqrt {2A}){e^{i{k_y}y}}{\phi _\lambda }\left( x \right)$, 
with sample area $A$ and the conserved parallel component of the wave vector $k_y$ in the ballistic transport. 
Right from the barrier, ${\phi _\lambda }\left( x \right)$ is a combination
of the two plane waves with transmission coefficient, $t_{\lambda \lambda}$ and $t_{\lambda \bar \lambda}$, where $\bar \lambda= -\lambda$, to describe 
intraband and interband scattering processes. 

By matching  ${\phi _\lambda }\left( x \right)$  and $d{\phi _\lambda }\left( x \right)/dx$  in the regions, $x<-d/2$,  $-d/2<x<-d/2$,
and $x>d/2$, we  obtain  $t_{\lambda \lambda}$, $t_{\lambda \bar \lambda}$, and 
the particle current density of the $\lambda$ channel
\begin{equation}
{j_\lambda }{\text{ = }}\operatorname{Re} \left[ v/A{\left( {{{\left| {{t_{\lambda \lambda }}} \right|}^2}\cos {\varphi _\lambda } + 
{{\left| {{t_{\lambda \bar \lambda }}} \right|}^2}\cos{\varphi_{\bar \lambda }}} \right)} \right],
\label{eq:current2}
\end{equation}
here, the group velocity of the scattered particle, $v = \sqrt {( {\alpha/\hbar)^2 + 2E/m^\ast}}$, has the same magnitude for the two bands. 
This current contains contributions from
the intra- and inter-channel transmission, where  $\varphi_\lambda$ is the incident and $\varphi_{\bar{\lambda}}$ the propagation angle of the cross-channel wave with the conservation of the $k_y$ component and 
$\varphi _{\bar{\lambda}}$ is related to $\varphi _{ \lambda }$ by $\cos {\varphi _{\bar{\lambda}}} = {\sqrt {k_{\bar{\lambda}}^2 - k_\lambda ^2{{\sin }^2}\varphi_{\lambda} } }/{k_{ \bar{\lambda}}}$, where 
\begin{equation}
k_\lambda = \sqrt {{{\left( {\frac{{\alpha m^\ast}}{{{\hbar ^2}}}} \right)}^2} + \frac{{2m^\ast E}}{{{\hbar ^2}}}}- \frac{{\lambda \alpha m^\ast}}{{{\hbar ^2}}}. 
\label{eq:kE}
\end{equation}

In the low-bias limit, i.e. $\left| {eV} \right| \ll {E_F}$, we get the expression for the conductance $G_\lambda$ in the $\lambda$ channel~\cite{Girvin:2019}
\begin{equation}
{G_\lambda } = \frac{e^2}{h}\frac{D}{2\pi}\int_{ - \frac{\pi }{2}}^{\frac{\pi }{2}} {d{\varphi _\lambda }k_F^\lambda {T_\lambda }\cos {\varphi_\lambda }},
\label{eq:conductance1}
\end{equation}
where  $D$ is the sample width, $k_F^\lambda$ the  $\lambda$-channel Fermi wave vector [$E=E_F$ in Eq.~(\ref{eq:kE})]
and the transmission is 
\begin{equation}
{T_\lambda } = \operatorname{Re} \left[{\left| {t_{\lambda \lambda}} \right|^2} + {\left| {t_{\lambda \bar \lambda }} \right|^2}\left( {\cos {\varphi _{\bar \lambda }}/\cos {\varphi _\lambda }} \right)\right].
\label{eq: transmission}
\end{equation}
The total conductance is the sum of the two channels, 
\begin{equation}
{G} = \sum\limits_{\lambda  =  \pm 1} {G_\lambda }.
\label{eq:G2DEG}
\end{equation}
To explore the evolution of the conductance as a function of the proximity-induced field and its direction, 
we consider its normalized value in Eq.~(\ref{eq:G2DEG}) expressed in terms of the Sharvin conductance for the 2DEG system
\begin{equation}
G_0^\text{2D} = \frac{e^2}{h}\frac{2D}{\pi}\sqrt{{{\left( {\frac{\alpha \, m^\ast}{\hbar ^2}} \right)}^2} + \frac{2m^\ast \,E_F}{\hbar ^2}.}
\label{eq:Sharvin}
\end{equation}

As shown in Fig.~\ref{fig:setup}(c), the $\mathcal{P}\sigma_z\mathcal{T}$ symmetry leads to perfect transmission 
and influences the conductance for the 2DEG system. 
Specifically, scattering states which satisfy $\mathcal{P}\sigma_z\mathcal{T}\psi(x,y)=\xi\psi(x,y)$, with eigenvalues of the form $\xi=e^{i\eta}$,
include the perfect transmission either due to effectively vanishing barrier or the resonances expected from the standing-wave condition.
However, $\mathcal{P}\sigma_z\mathcal{T}$ symmetry 
is not an intrinsic symmetry of the system and $\mathcal{P}\sigma_z\mathcal{T}$ does not commute with $H_\text{F/2DEG}$ in Eq.~(\ref{eq:H2DEG}).  
Instead, the $\mathcal{P}\sigma_z \mathcal{T}$ symmetry emerges only for certain specific system parameters and scattering states satisfying,
\begin{equation}
[H_\text{F/2DEG},\mathcal{P}\sigma_z\mathcal{T}]\psi_R(x,y)=0,
\label{eq:commutator}
\end{equation}
where the index $R$ emphasizes that the relation is restricted only to 
the cases of the perfect transmission (for vanishing of the effective barrier or
at resonances). This symmetry generalizes a simple case of the perfect transmission 
in a potential barrier (or a spinless) system~\cite{Griffiths:2005}.

\begin{figure}[t]
\centering
\includegraphics*[trim=0.4cm 0.4cm 0.1cm 0.6cm,clip,width=8cm]{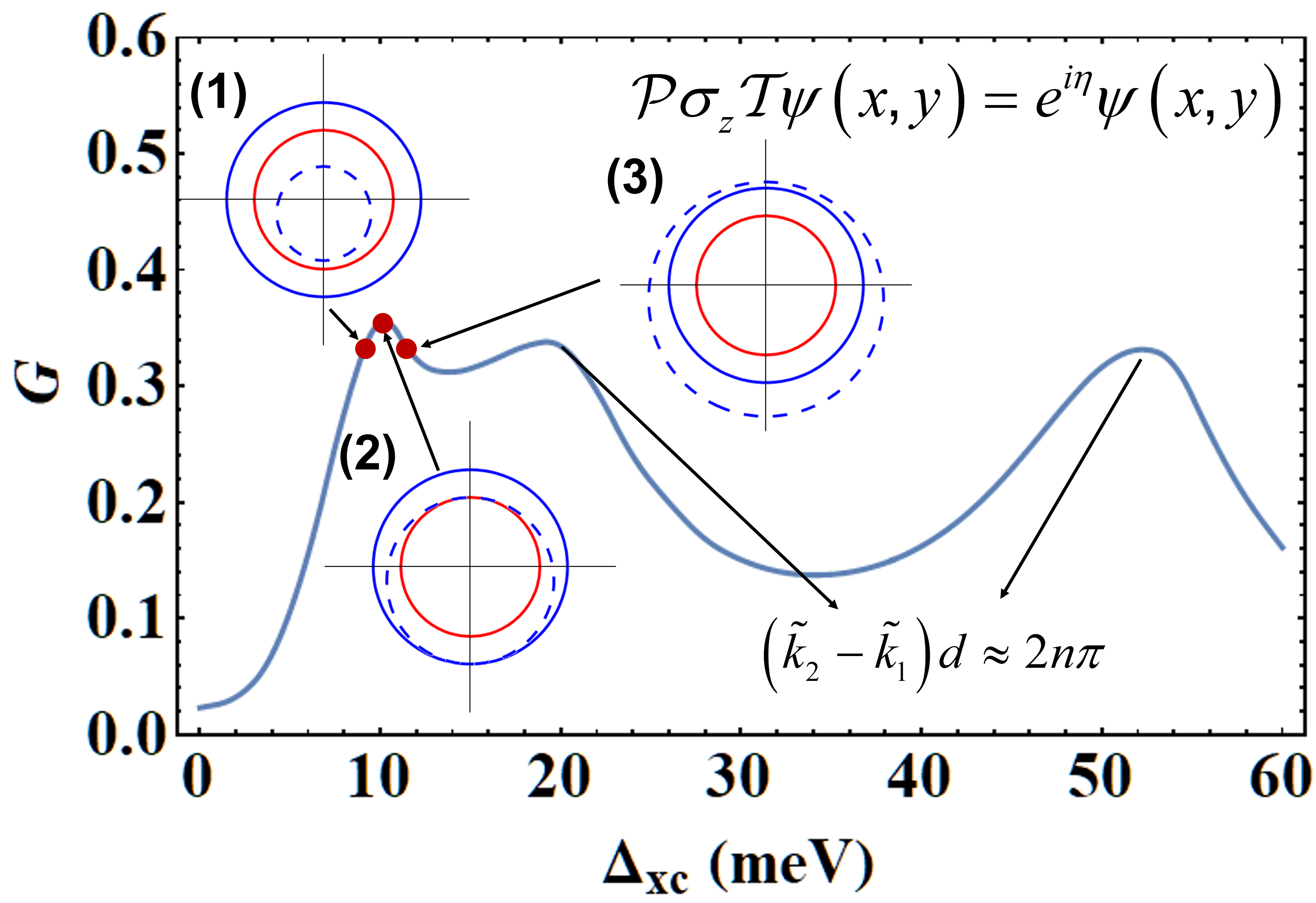}
\caption{
Normalized conductance, $G$, as a function of $\Delta_\text{xc}$, for $E_F = 5\,$meV, $V_0=10\,$meV, 
$\alpha=0.093\,$eV\AA, and $d=26\,$nm.
Fermi contours at different $\Delta_\text{xc}$, ${\bf m} || {\bf x}$: Blue and red circles are in the lead and the dashed contours in the barrier (only one band exists at the given $E$).
Each of the three peaks in $G$ satisfies the $\mathcal{P}\sigma_z\mathcal{T}$ symmetry for the perfect transmission.} 
\label{fig:contour}
\end{figure}

To understand the relation between conductance, barrier parameters, and the perfect transmission, 
we perform a Fermi contour analysis illustrated in Fig.~\ref{fig:contour}.
In the spinless case, with the better matching between Fermi contours in the lead and barrier a higher 
transmission is expected.  
For a simple 1D geometry (normal incidence or a fixed $k_y$) and $\delta$-barrier, the transmission is~\cite{Blonder1982:PRB} 
\begin{equation}
T{\text{ = 1/}}\left( {1 + {Z^2}} \right), 
\label{eq:Z}
\end{equation}
where $Z$ is the effective barrier strength which combines the influence of a native barrier and the Fermi wave vectors mismatch in the two regions~\cite{Zutic1999:PRBa,Vezin2020:PRB,Blonder1983:PRB,Zutic2000:PRB}. 
Since the difference between the lead and barrier Fermi contours is associated with the effective 
barrier potential, a larger mismatch between Fermi contours corresponds to a larger effective $Z$ and thus to a lower transmission.

We now generalize this situation to a spin-dependent case in the presence of  $\Delta_\text{xc}$ and $\alpha$, as depicted in Figs.~\ref{fig:bands}(b) and (c). 
The effective barrier at the F/2DEG interface becomes spin-dependent since the energy bands in the barrier are
split by the exchange field and spinless $Z$ becomes $Z_\pm^\text{eff}$~\cite{Vezin2020:PRB}. 
For the situation from Fig.~\ref{fig:bands}, we can decompose the incident spinor in the basis constructed by the corresponding barrier eigenspinor  ($k_{x,y}$ is
the same as in the incident state)
for the lower band $\chi_\uparrow$ and its antiparallel 
partner from the upper band $\chi_\downarrow$, i.e., $\left | \chi_\text{in} \right\rangle  =  \left\langle \chi_\uparrow  {\left |\right.}\chi_\text{in} \right\rangle 
\left | \chi _\uparrow \right\rangle+\left\langle \chi _\downarrow  {\left |\right.} \chi _\text{in} \right\rangle \left | \chi _\downarrow \right\rangle$.
The first term undergoes a weak effective barrier $Z_-^\text{eff}$ while the second term experiences a strong barrrier $Z_+^\text{eff}$.

The effective interfacial barrier is inequivalent for two helicities (for outer/inner Fermi contours) leading to an
important influence of spin mismatch on transmission, not just the mismatch of the Fermi contours~\cite{Vezin2020:PRB}.
Considering that the direction of
the proximity-induced exchange field, ${\bf m}$ [see Eq.~(\ref{eq:H2DEG})] is determined by $\phi$, for an in-plane ${\mathbf M}$ as depicted in  Fig.~\ref{fig:setup}(c), 
the effective barrier for $\left | \chi_\text{in} \right\rangle$ from the outer band of the lead is
\begin{equation}
Z_ \pm^\text{eff}  \propto {V_0} \pm \sqrt {{{\left( {\alpha {k_x} + {\Delta _\text{xc}}\sin \phi } \right)}^2} + {{\left( {\alpha {k_y} - {\Delta _\text{xc}}\cos \phi } \right)}^2}}  - \alpha k.
\label{eq:Zout}
\end{equation}
Equivalently, for $\left | \chi_\text{in} \right\rangle$ from the inner band of the lead 
\begin{equation}
Z_ \pm^\text{eff}  \propto {V_0} \pm \sqrt {{{\left( {\alpha {k_x} + {\Delta _\text{xc}}\sin \phi } \right)}^2} + {{\left( {\alpha {k_y} - {\Delta _\text{xc}}\cos \phi } \right)}^2}}  + \alpha k.
\label{eq:Zin}
\end{equation}
In Eqs.~(\ref{eq:Zout}) and (\ref{eq:Zin}) 
the impact of the reflected states is neglected, and this treatment is most accurate for small effective $Z$, i.e.,  for a good matching of Fermi contours.
We note that the effective barrier becomes energy-dependent when  
$\Delta_{xc} \neq 0$ and $\alpha \neq 0$, 
opening the possibility for resonant tunneling to occur when the energy of the incident carrier is such that $Z_-^\text{eff}\rightarrow 0$   (or $Z_+^\text{eff}\rightarrow 0$).

Due to the spin mismatch, we need to include a correction of ${\left| {\left\langle {{\chi _ \uparrow }|{\chi _{{\text{in}}}}} \right\rangle } \right|^2}$ in the transmission, i.e.,
$T \approx {\left| {\left\langle {{\chi _ \uparrow }|{\chi _{{\text{in}}}}} \right\rangle } \right|^2} T_{\chi_\uparrow}$, where 
$T_{\chi_\uparrow}$ is the transmission without the spin mismatch. 
We can see from Fig.~\ref{fig:setup}(c) that the spin mismatch is much smaller for the incident state from the outer lead band $(\lambda=-1)$, i.e.,
$\left| {\left\langle {{\chi _ \uparrow }} \mathrel{\left | {\vphantom {{\chi _ \uparrow } {\chi _{\text{in}},\lambda  =  - 1}}}	\right. \kern-\nulldelimiterspace} {{\chi _{\text{in}}},\lambda  =  - 1} \right\rangle } \right| \gg \left| {\left\langle {{{\chi _ \uparrow }}}	\mathrel{\left | {\vphantom {{\chi _ \uparrow } {{\chi _{\text{in}}},\lambda  = 1}}} \right. \kern-\nulldelimiterspace}{{\chi _{\text{in}}},\lambda  = 1} \right\rangle } \right|$,
which indicates that the most of $G$ is contributed by the incident states from the outer lead band.
When the state inside the barrier is the same (up to a phase) as that in the lead, perfect transmission is achieved.

With vanishing Rashba SOC, $\alpha \rightarrow 0$, outer and inner lead Fermi contours coincide and the spin-dependent barrier, 
$V_\pm=V_0\pm \Delta _\text{xc}$, from Fig.~\ref{fig:bands}(a) is recovered. We focus on $V_0>0$, $V_-=0$ leads to the perfect transmission,
$T_{\chi_\uparrow}=1$, from our discussion above. From Eqs.~(\ref{eq:Zout}) and (\ref{eq:Zin}), 
$Z_\pm^\text{eff} \propto V_0 \pm \Delta_\text{xc}$, so one can still get $Z_-^\text{eff}\rightarrow 0$ when $V_0 = \Delta_\text{xc}$.
We will further address the importance of vanishing $V_-$ in F/S junctions.
At a fixed $k_y$, the eigenstate for the perfect transmission $T(k_y)=1$, 
precisely satisfies the $\mathcal{P}\sigma_z\mathcal{T}$ symmetry: 
 $\mathcal{P}\sigma_z\mathcal{T}\psi(x,y)=\xi\psi(x,y)$, with eigenvalue $\xi=e^{i\eta}$ and an arbitrary phase $\eta$.

This analysis and the relevance of  the $\mathcal{P}\sigma_z\mathcal{T}$ symmetry can be applied to Fig.~\ref{fig:contour} with
parameters of a typical InGaAs/InAlAs 2DEG, $m^\ast = 0.05 m_0$, where
$m_0$ is the free-electron mass, and $\alpha = 0.093\,$eV\AA~\cite{Nikolic2005:PRB}.
It is crucial to note that $\mathcal{P}\sigma_z\mathcal{T}$ symmetry arguments
for a perfect transmission pertain to a single $k_y$ (single channel), 
while the geometry from Figs.~\ref{fig:setup}(a) and~\ref{fig:bands} corresponds to a 2D system where
the total $G$ reflects the contribution from all allowed $k_y$. In the spinless case, maximizing $G$ means
an overall contour matching, while even with adding spin our Fermi contour analysis provide a valuable
tool to visualize and understand various trends in $G$. 

With conserved $k_y$, from Fig.~\ref{fig:bands}(c) 
we can recognize many similarities with the Snell's law~\cite{Zutic2000:PRB}. We may get propagating or 
decaying (evanescent) states in different regions, depending on the incident angle, $\varphi$ and the particular 
contour (outer, inner) considered. For example, for carriers from the $\lambda=-1$ band (outer contour) and 
incident angles $\varphi$ exceeding the critical angle,
\begin{equation}
\varphi _c =  \pm \arcsin(k_1/k_{-1}), 
\label{eq:critical}
\end{equation}
where $k_\lambda=\pm1$ is given by
Eq.~(\ref{eq:kE}), the transmission and reflection 
to the $\lambda=1$ band are not allowed because there are no such propagating states. In this regime, 
back scattering is suppressed while $T_{\lambda=-1} $  is enhanced.

The overall maximum in $G(\Delta _\text{xc})$ is related to the perfect transmission
due to the Fermi contour matching depicted by the example (2) in Fig.~\ref{fig:contour}. As shown for examples (1) and (3)
a small change in $\Delta _\text{xc}$ leads to worse Fermi contour matching and reduced $G$.
The origin of the overall peak in $G$ near $\Delta _\text{xc}=V_0$ can be further understood in the small SOC limit.
The barrier contour reduces to a shifted circle with radius $k_F\sqrt{\left( {{E_F} + \Delta  - {V_0}} \right)/{E_F}}$, 
where we recall that $k_F$ is average of the outer and inner lead circles. The barrier contour can then be approximated by
\begin{equation}
	\begin{gathered}
		{\left[ {k_x - (m^\ast \alpha/\hbar ^2)\sin \phi } \right]^2} + {\left[ {k_y + (m^\ast \alpha/\hbar ^2)\cos \phi } \right]^2} \hfill \\
		\hspace*{90pt} \approx (2m^\ast/\hbar ^2)\left( {E_F + \Delta  - V_0} \right). \hfill \\
	\end{gathered} 
	\label{eq:barrier_contour}
\end{equation} 

In the region near the $\Delta_\text{xc}=V_0$, both $G(\phi=0)$ and $G(\phi=\pi/2)$, which 
correspond to ${\bf m}\parallel{\bf x}$ and ${\bf m}\parallel{\bf y}$, reach their maxima 
because of the best matching of Fermi contours between lead and barrier. 
The shift  of barrier circle is always $\perp$ to $\mathbf{m}$ and is of the first order in $\alpha$, while the deformation is a higher-order correction.

Remarkably, the other local maxima in $G$, for larger $\Delta _\text{xc}$ in Fig.~\ref{fig:contour}, also satisfy the $\mathcal{P}\sigma_z\mathcal{T}$ symmetry. 
However, instead of the  
contour matching, they are due to standing de Broglie waves in the barrier and the constructive interference.
With outer/inner lead contours, there are four eigenstates in the leads with $\pm k_{x \lambda}, \pm k_{x \tilde{\lambda}}$, we can
distinguish cases: (i) 4 (4), (ii) 4 (2), (iii) 2 (4), and (iv) 2 (2) propagating states in the leads (barrier)~\cite{Shen2020:PRB}. 

In cases (i)  and (ii) [arising for an incident state from the inner band of for the outer band and  $\varphi<\varphi_c$], 
the perfect transmission means that no reflected wave should exist in the left lead, while the state in the right lead should match,
 up to a phase, the corresponding one in the left lead. For case (i), 
under $\mathcal{P}{\sigma _z}\mathcal{T}$ symmetry operation, up to a phase, all the propagating states inside the barrier,
with $x$ component of the wave vector, $\tilde{k_j}$, j=1,...,4, remain the same.
By matching  ${\phi _\lambda }\left( x \right)$  and $d{\phi _\lambda }\left( x \right)/dx$, as noted in Sec.~II, the resonance condition is satisfied when
\begin{equation}
\left( {{{\widetilde k}_j} - {{\widetilde k}_1}} \right)d = 2\pi{n_j},\quad j=2,3,4,
\label{eq:case1}
\end{equation}
where $n_j=1,2,...$ However, for the system we consider, ${{\widetilde k}_j}d \sim 10$ and
Eq.~\eqref{eq:case1} only holds in few special cases.  

In case (ii) the resonance condition similar to Eq.~(\ref{eq:case1}) is not possible,
unless for the normal incidence when $\mathbf{m}\parallel\mathbf{y}$. 
In that case, the coefficients of the decaying barrier states vanish and the spins of all scattering states become parallel to each other
which makes the system ``spinless" and perfect transmission becomes possible.

In cases (iii) and (iv), the presence of the decaying states in the leads can be understood for the incident particle from the $\lambda=-1$ band
at an angle $\varphi>\varphi_c$. Since such decaying states in the leads do not carry any current~\cite{Zutic2000:PRB}, as can be understood 
from the Snell's law, the perfect transmission and the resonance conditions are still possible. 
In case (iv), under the  symmetry $\mathcal{P}{\sigma _z}\mathcal{T}$ operation, 
up to a phase, the two propagating states inside the barrier remain the same, while the two decaying states in the barrier become, up to a phase,
the two decaying states in the leads. The resonance condition is
\begin{equation}
\left({\tilde k}_2-{\tilde k}_1\right)d = 2n\pi-\delta, \quad n=1,2,3\ldots, 
\label{eq:case2}
\end{equation}
where for the $x$ component of the propagating wave vectors in the barrier we assume  ${\tilde k}_2>{\tilde k}_1$,  while
 the correction $\delta \propto (\alpha/\hbar)\sqrt {m^\ast /\Delta _\text{xc}} \ll 1$. 
Since the resonances occur at ${\Delta_\text{xc}}> E_F$,
the magnitude of the propagating barrier wave vector is much larger than $k_y$. Therefore, the $x$ component 
of the propagating barrier wave vector is almost the same for all the incoming states with different incident angles, which means the transmission
resonances occur for all the incoming states almost simultaneously when the resonance condition is satisfied and thus the maximum $G$ is reached.

The two local maxima in $G$ near $\Delta _\text{xc} =20\,$meV and 55$\,$meV, shown in Fig.~\ref{fig:contour}, 
both correspond to the case (iv) and a large $\Delta _\text{xc}$ regime,
where for $\mathbf{m}\parallel\mathbf{x}$ the propagating  wave vectors are ${{\tilde k}_{1,2}} \approx  \mp \sqrt {2m^\ast \Delta}/\hbar$.
Applying the resonance condition from Eq.~(\ref{eq:case2}) the first (second) of these maxima corresponds to $n=1$ ($n=2$). However,
with large $\Delta _\text{xc}$, the correction $\delta$ in Eq.~(\ref{eq:case2}) is very small such that the resonance condition is accurately described
by the spinless case where $\delta=0$.

\begin{figure*}[t]
\centering
\includegraphics*[trim=0.4cm 0cm 0cm 0cm,clip,width=15cm]{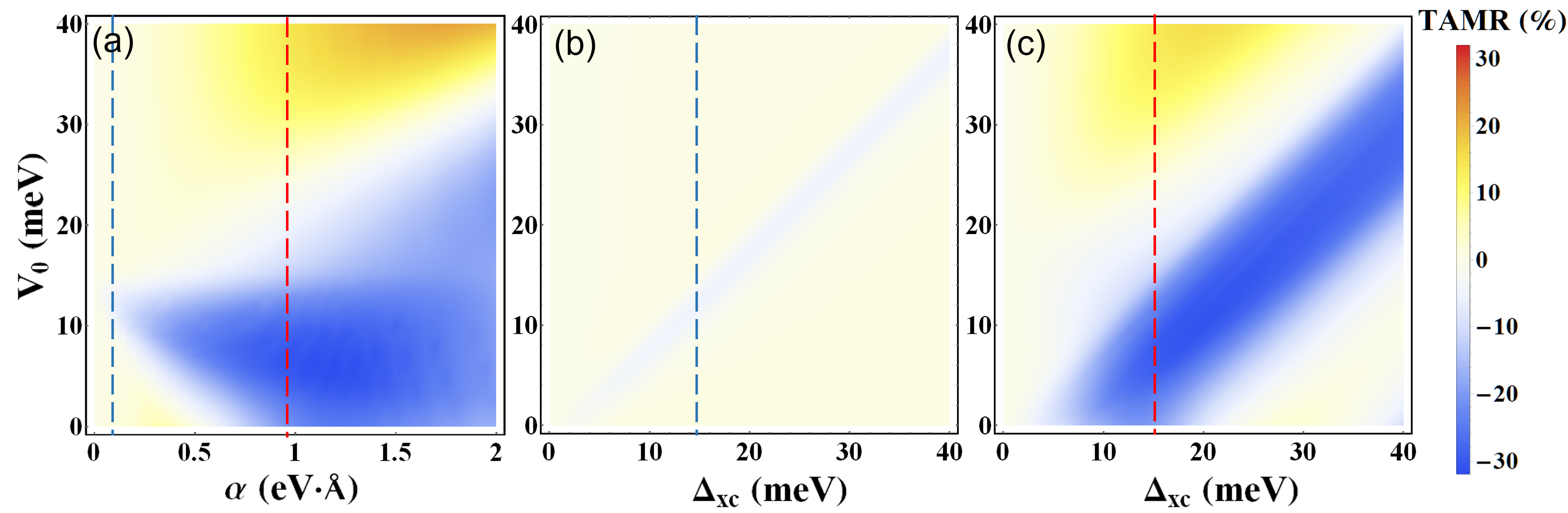}
\caption{(a) Dependence of the TAMR amplitude on the Rashba SOC $\alpha$ and the barrier strength $V_0$ for a 2DEG system with a $d = 13\,$nm thick barrier, $E_F = 10\,$meV, 
and proximity-induced exchange splitting $\Delta_\text{xc}=15\,$meV. 
(b) Dependence of the TAMR amplitude on $\Delta_\text{xc}$ and  $V_0$, for the same $d$ and $E_F$ as in (a), with $\alpha=0.093\,$eV$\text{\AA}$ for a typical InGaAs/InAlAs 2DEG.
(c) The same as (b), but  $\alpha=0.93\,$eV$\text{\AA}$.
The blue and red dashed lines in (a) are for the same parameters as the ones in (b) and (c).
}
\label{fig:TAMR}
\end{figure*}

\section{III. F/2DEG In-Plane TAMR}

A large value of TAMR, as defined in Eq.~(\ref{eq:tamrIN}), has the origin in the difference for the maximum $G(\phi=0)$  and $G(\phi=\pi/2)$, that is
for the $\mathbf{m}\parallel\mathbf{x}$ and $\mathbf{m}\parallel\mathbf{y}$ orientations.  In Fig.~(\ref{fig:contour}), for $G(\phi=0)$ and shown
examples (1)-(3), we see the expected barrier Fermi contour (broken line) shifted downward ($\perp \mathbf{x}$). This broken $k_y$ symmetry leads
to an unusual tunneling planar Hall effect, where the transverse voltage is maximum for the $\mathbf{M}\parallel\mathbf{x}$, while other Hall effects 
would vanish~\cite{Scharf2016:PRL}. 
Due to the same asymmetry, the perfect match for both upper and lower half of the circle cannot be achieved simultaneously. 
We see in examples (1)-(3) that the upper (lower) half of the barrier contour tends to match the inner (outer) lead contour 
since the spin mismatch is smaller for these states. 
The best possible simultaneous match of Fermi contours yields the large transmission at $\Delta_\text{xc}=V_0$, resulting
in the first peak in $G$, even for large $\alpha$ when the barrier contour is strongly 
deformed from a circle, as in Fig.~\ref{fig:bands}(c).  

As we  see from Fig.~\ref{fig:contour} with multiple peaks in $G$, 
there also exist  several resonances which could influence the TAMR. 
In the limit of $\alpha \to 0$, the condition for such resonances can be
derived from Eq.~(\ref{eq:case2}) with $\delta=0$
\begin{equation}
\Delta_\text{xc}  = {V_0} - {E_F} + \frac{\pi ^2 \hbar ^2 n^2}{2m^\ast d ^2},
\label{eq:Delta_max}
\end{equation}
for both $G(\phi=0)$ and $G(\phi=\pi/2)$ which correspond to ${\bf m}\parallel{\bf x}$ and ${\bf m}\parallel{\bf y}$,
where $n=1,2,...$ describes the order of the resonance that pertains to the case (iv) in Sec.~II. In this situation, 
of vanishing Rashba SOC, the conductance no longer depends on the in-plane ${\bf m}$ and we expect a vanishing TAMR, as inferred from Eq.~(\ref{eq:tamrIN}).

However, already for a small and nonvanishing $\alpha$ the maximum $G(\phi=\pi/2)$ is achieved for a different condition when the barrier circle 
shares the same size as the outer circle of the lead~\cite{Shen2020:PRB}. Up to the first order in $\alpha$ the condition leading to the 
maximum $G(\phi=\pi/2)$ is given by $\Delta_\text{xc}  = V_0 + \alpha k_F$. 
When such a condition is satisfied, the barrier Fermi contour matches the outer lead contour instead of the inner one, because the main $G$ contribution 
is from the incident particles on the outer Fermi contour. Correspondingly, near $\Delta_\text{xc}  = {V_0}$, 
due to contour matching and the perfect transmission, we expect enhanced TAMR, as can be seen from Fig.~\ref{fig:TAMR}.
 
Furthermore, when $\alpha \ne0$, ${{\tilde k}_2 - {\tilde k}_1}$, which determine the resonance condition in Eq.~(\ref{eq:case2}),
varies as   ${\mathbf M}$ is rotated. This means the maxima for $G(\phi=0)$ and $G(\phi=\pi/2)$ will be achieved at different $\Delta_\text{xc}$. 
Therefore, similar to the situation near $\Delta_\text{xc}=V_0$, an enhanced 
TAMR will arise from the small difference in the peak conditions. 
With fixed $E_F$ and $V_0$, up to the lowest-order correction of $\alpha$, we can derive the difference in $\Delta_\text{xc}$ at the same 
order maxima ($n$) of $G(\phi=0)$ and $G(\phi=\pi/2)$
\begin{equation}
\Delta^\text{diff}_\text{xc} \propto \alpha ^2m^\ast/\hbar^2.
\label{higher order resonance peak shift}
\end{equation}
Unlike the perfect transmission condition due to contour matching, the difference in now quadratic in $\alpha$. 

With this conductance analysis and nonmonotonic $G(\Delta_\text{xc})$ in Fig.~\ref{fig:contour}, a consequence of the collective 
contributions of multiple $T\approx 1$ 
states corresponding to different propagation directions of the tunneling carriers, we expect
various nonmonotonic TAMR trends and a strong influence of Rashba SOC. We can see from Fig.~\ref{fig:TAMR}(a) 
that increasing $\alpha$ can enhance both TAMR and the range of $V_0$ for such enhancement. 
However, if $\alpha$ gets too large ($>1.5\,$eV\AA), the absolute value of TAMR decreases, for a large range of $V_0$.

The reason for this unexpected trend in TAMR($\alpha$) can be understood from calculated results in Figs.~(\ref{fig:TAMR})(b) and (c), 
which are shown for two different values of $\alpha$, as indicated by blue and red vertical lines in Fig.~(\ref{fig:TAMR})(a).
From the previous discussion  of a small $\alpha$ limit, we know that $G$ peaks near $\Delta_\text{xc}=V_0$ and such a peak 
will be shifted by $\alpha k_F$ when ${\bf M}$ is rotated. The sensitive dependence of $G$ on the ${\bf M}$ orientation leads 
 to a large TAMR near the diagonal line in the parameter space, shown in Figs.~(\ref{fig:TAMR})(b) and (c). 
 Since such a shift is proportional to $\alpha$, by comparing Figs.~(\ref{fig:TAMR})(b) and (c) there is a wider range of the enhanced TAMR for a larger $\alpha$. 
 
 However, if $\alpha$ is too large (compared to $V_0/k_F$ and $\Delta_\text{xc}/k_F$), the impact of the ${\bf M}$ orientation becomes negligible, and 
 TAMR starts to decrease. This can be seen at the lower-left corner in Fig.~(\ref{fig:TAMR}(c), where the resonant TAMR is much smaller when both 
 $V_0/k_F$, $\Delta_\text{xc}/k_F<\alpha$.  Even for a fixed $V_0$, for example at 5$\,$meV, we see that with $\alpha$ 
 the absolute value of TAMR is nonmonotonic and TAMR also changes sign.
 
 The calculated in-plane TAMR reveals various other peculiar trends. For a fixed $\alpha$ [Fig.~(\ref{fig:TAMR}(a)] or $\Delta_\text{xc}/k_F$ [Fig.~(\ref{fig:TAMR}(b) and (c)],
 we can see that TAMR can be nonmonotonic in $V_0$ and even change its sign. 
 Furthermore, there is as a clear nonmonotonic amplitude with $\Delta_\text{xc}$ at a fixed $V_0$ in 
 Fig.~(\ref{fig:TAMR}(b). Different barrier thickness in Fig.~\ref{fig:TAMR}  (26$\,$nm) from that in Fig.~\ref{fig:contour} (13$\,$nm) modify the respective 
 $\Delta_\text{xc}$ 
 values for the perfect transmission.  These different TAMR trends in Fig.~\ref{fig:TAMR} were primarily related to the best contour matching, near $\Delta_\text{xc}=V_0$ 
 [as the first peak in $G$, seen in Fig.~\ref{fig:contour})], while the role of $n=1$ resonance can be seen in the lower-right corner in Fig.~\ref{fig:TAMR}(c).

From the angular dependence of TAMR, it is possible to obtain valuable information about the interfacial crystallographic symmetry~\cite{Zutic2019:MT}.
For a 2DEG system, in the limit 
and $\Delta_\text{xc}/V_0 \ll 1$,  the leading contribution to  the in-plane angular dependence of $G$ from the two incoming channels with helicity $\lambda=\pm1$,  
is  $\pm \sin \phi$.  However, with their opposite signs, these leading contributions cancel in the total  $G$,
which becomes significantly smaller, quadratic in the small parameter, and has a different angular dependence, 
resulting in 
\begin{equation}
\text{TAMR}_{\text{F/2DEG}} \left( \phi  \right) \sim (\Delta_\text{xc} /V_0)^2 \sin^2 \phi.
\label{eq:TAMR_2DEG}
\end{equation}
While this angular dependence also coincides with the results from the surface states of 3D topological insulators, their junctions with ferromagnets
do not have a resonant TAMR behavior of a 2DEG and thus lead to different trends in $\Delta_\text{xc}$ and $V_0$~\cite{Shen2020:PRB}.

Another striking signature of the underlying resonant behavior of the calculated TAMR is its magnitude. 
In commonly considered vertical tunneling devices, the in-plane TAMR is rather small (typically $< 1$ \%) even for large carrier spin polarization~\cite{Gould2004:PRL} and 
exchange energies $>$ eV~\cite{Moser2007:PRL,Park2008:PRL,Uemura2009:APL}. In contrast, with much smaller proximity-induced exchange fields, 
$\Delta_\text{xc}\sim 10\,$meV, we find a much larger TAMR in our lateral (planar) structures, $>10\,$\% even for SOC for a typical InAs-based 2DEG.

As we next focus on the F/S junctions and their superconducting out-of-plane TAMR analog [recall Eq.~(\ref{eq:tamrOUT})], we will see that some
of the calculated nonomonotonic trends, similar to those we discussed for F/2DEG systems, are experimentally verified, while the 
analysis of effective barriers allows us to give simple estimates of the enhanced MR which can even reach 100$\,$\% in the measured samples.  \\
 
\section{IV. F/S HAMILTONIAN AND CONDUCTANCE}

Motivated by the recent experiments demonstrating a large MR enhancement in F/S junctions~\cite{Cai2021:NC}, 
depicted in Fig.~\ref{fig:FS},  we consider a  simple model for F/S Hamiltonian, $H_\text{F/S}$.  It shares several similarities 
with $H_\text{F/2DEG}$ in Eq.~(\ref{eq:H2DEG}), 
used to analyze in-plane TAMR. The key difference in the superconducting state is the need to simultaneously include the presence of 
electrons and holes and the pair potential $\Delta$, which couples electronlike and holelike components of the underlying wavefunction.  

Additionally, while Rashba SOC was inherent for whole 2DEG region, in F/S junction it is only an interfacial contribution, 
along with the potential barrier, between F and S region. Now we study vertical transport and 
out-of-plane MR [recall Eq.~(\ref{eq:tamrOUT})], which is dominated by the  process of Andreev reflection, introduced in Sec.~I and 
it is the microscopic mechanism for the superconducting proximity effects~\cite{Zutic2019:MT}.

Before specifying $H_\text{F/S}$ and the resulting equation for quasiparticle states, it is instructive to note a similarity 
between the two-component transport in normal metal (N)/S junctions (for electronlike and holelike quasiparticles) and F/N junctions (for spin $\uparrow$, $\downarrow$), 
which both lead to current conversion, accompanied by additional boundary resistance~\cite{Zutic2004:RMP}. 
In the N/S junction Andreev reflection is 
responsible for the conversion between the normal and the supercurrent, characterized by the superconducting coherence length, while in the F/N case 
a conversion between spin-polarized and unpolarized current is characterized by the spin diffusion length. In the N/S charge transport,  
$\Delta$, which couples electronlike and holelike states, plays a similar role of the spin-flip potential which 
couples $\uparrow$, $\downarrow$ states in the F/N spin transport. 

We consider ballistic F/S junction, depicted in Fig.~\ref{fig:FS},  
having a flat interface at $z=0$ with potential and Rashba SOC scattering. 
Similar to the approach in Sec.~II of matching the wavefunctions for the scattering states in different regions,
we generalize the Griffin-Demers-Blonder-Tinkham-Klapwijk formalism\cite{Griffin1971:PRB, Blonder1982:PRB,Kashiwaya2000:RPP,Wu2009:PRB} 
to solve Bogoliubov-de Gennes equation for quasiparticle states $H_\text{F/S} \Psi \left( \bm{r} \right)= E \Psi \left( \bm{r} \right)$,
with energy $E$~\cite{Hogl2015:PRL}, where 
\begin{equation} 
H_\text{F/S}=\left( {\begin{array}{*{20}{c}}
	{{H_e}}&{\Delta\Theta(z) {I_{2 \times 2}}} \\ 
	{{\Delta ^*}\Theta(z){I_{2 \times 2}}}&{{ H_h}} 
	\end{array}} \right),
	\label{eq:H}
\end{equation}
with $\Delta$ the $s$-wave pair potential which, for a homogeneous S region, is also the value
of the superconducting gap, and  the single-particle Hamiltonian for electrons  is
\begin{equation}
\begin{aligned}
H_e=  - \frac{\hbar ^2}{2}\bm{\nabla} {\frac{1}{m(z)}}\bm{\nabla}  - \mu(z) - \Delta _\text{xc}(\bm{m} \cdot {\bm \sigma}) \Theta(- z) + H_B,
\label{eq:He}
\end{aligned}
\end{equation}
while for holes $H_h=-\sigma_y H_e^*\sigma_y$.
In Eq.~(\ref{eq:H}) $m(z)$ is the effective mass and  $\mu(z)$ is the chemical potential.
The interfacial scattering is modeled by delta-like potential barrier, 
with the effective barrier height $V_0$ and width $d$ and the interfacial Rashba SOC of
strength $\alpha$ has different units than Rashba SOC in F/2DEG~\cite{Moser2007:PRL} 
\begin{equation}
\begin{aligned}
H_B=[V_0 d + \alpha (k_y {\hat \sigma }_x - k_x{\bf \sigma}_y)]\delta(z).
\label{eq:HB}
\end{aligned}
\end{equation}
As in Sec.~II,  $\Delta_{xc}$ is the exchange spin splitting
and  we denote the  orientation of the magnetization, $\bf{M}$, 
by $\bf{m}$, but we now also consider its out-of-plane rotation
$\bf{m} = \left( {\sin \theta \cos  \phi , \sin \theta \sin \phi , \cos \theta } \right)$. 

Since the in-plane wave vector $\bm{k}_\parallel$ is conserved, the scattering states for incident spin $\sigma$ electron are given by 
$\Psi_\sigma \left(\bm{r} \right) = e^{i\bm{k}_\parallel \cdot \bm{r}_\parallel}   \psi_\sigma \left(z \right)$ in a four-component basis, 
they include Andreev and specular reflection, 
each without and with spin flip, given by the amplitudes $a_\sigma$, $b_\sigma$, $\bar{a}_\sigma$, and $\bar{b}_\sigma$~\cite{Zutic1999:PRBa}.

In the F region, the eigenspinors for electrons and holes are 
$\chi _\sigma^e = {\left( {{\chi _\sigma },0} \right)^T}$ and $\chi _\sigma ^h = {\left( {0,{\chi _{ - \sigma }}} \right)^T}$ with
\begin{equation}
{\chi _\sigma} = (1/\sqrt{2}){\left( {\sigma \sqrt {1 + \sigma \cos \theta } {e^{ - i\phi }},\sqrt{1 - \sigma \cos \theta }} \right)}^T,
\label{eq:spinor}
\end{equation}
where $\sigma=1 (-1)$ or $\uparrow$ ($\downarrow$) refer to spin parallel (antiparallel) to $\bf{M}$ and the $z$ components of the 
wave vector are
\begin{equation}
k^{e\,(h)}_\sigma = \sqrt{k_F^2+\frac{2m_F}{\hbar^2}\left[(-)E + \sigma \Delta_\text{xc}\right]-k_\parallel^2}, 
\label{eq:keh}
\end{equation}
with a spin-averaged
Fermi wave vector, $k_F$~\cite{Zutic2000:PRB}. 
At an interface,  with conserved ${\bm k}_\|$,  the  $H_B$ barrier eigenspinors 
in the helicity basis  are given  by~\cite{Fabian2007:APS,Matos-Abiague2009:PRB},
\begin{equation}
{\chi _ \pm } = \frac{1}{{\sqrt 2 }}\left( {\begin{array}{*{20}{c}}
	1 \\ 
	{ \mp i{e^{i\gamma }}} 
	\end{array}} \right),
\label{eq:Bspinor}
\end{equation}
where $\gamma$ is the angle between $\bm{k_\|}$ and $\hat{\bm{k}}_x$.

In the S region, coherence factors, $u$, $v$, satisfy
${u^2} = 1 - {v^2} = \left( {1 + \sqrt {{E^2} - {\Delta ^2}} /E} \right)/2$, while the $z$-components of the 
wave vector are
\begin{equation}
q^{e\,(h)} = \sqrt{q_F^2+(-)\frac{2m_S}{\hbar^2}\sqrt{E^2-\Delta^2}-k_\parallel^2}, 
\label{eq:qeh}
\end{equation}
with $q_F$ the Fermi wave vector. 
Similar to Snell's law~\cite{Zutic2000:PRB}, for a large $k_\parallel$ these $z$-components can become imaginary
representing evanescent states which carry no net current.
The vanishing of evanescent states at infinity requires 
$\operatorname{Im} [{k_\sigma ^{h}} ]  < 0$, so  the sign of the $z$-component of the wave vectors needs to be chosen correctly.

From the charge current conservation, the zero-temperature differential conductance at applied bias, $V$,
can be expressed as~\cite{Hogl2015:PRL}
\begin{equation}
G(V) =  \sum\limits_\sigma  \int \frac{d\bm{k}_\parallel}{2 \pi k_F^2} \left[ 1 + R_\sigma^h(- eV) - R_\sigma^e(eV) \right].
\label{eq:G}
\end{equation}
Here $G(V)$ is normalized by the Sharvin conductance~\cite{Zutic2004:RMP} 
\begin{equation}
G_0^{\text{3D}}= \frac{e^2 k_F^2 A}{2 \pi h},
\label{eq:Sharvin3}
\end{equation}
where $A$ is the interfacial area, while the form is different than 
the previous expression in 2D [recall Eq.~(\ref{eq:Sharvin})], 
where SOC is present in the whole 2D region, not just at an interface.
Only the probability amplitudes from the F region are needed,  
for  Andreev  $R_\sigma^h = \operatorname{Re} [(k_{-\sigma}^h/k_\sigma^e){\left| a_\sigma \right|^2} + (k_\sigma^h/ k_\sigma^e){\left| {\bar{a}_\sigma} \right|^2}]$ 
and specular reflection $R_\sigma^e = \operatorname{Re} [{\left| b_\sigma \right|^2 + (k_{-\sigma}^e/k_\sigma^e) \left| \bar{b}_\sigma \right|^2}]$.
A finite-temperature correction is straightforward by adding Fermi functions in Eq.~(\ref{eq:G}), which smear the calculated $G(V)$ at $T=0\,$K.
The integration kernel in Eq.~(\ref{eq:G}), $[1+ R_\sigma^h-R_\sigma^e]$ can be viewed as the effective transmission. 

\begin{figure}[t]
\vspace{-0.1cm}
\centering
\includegraphics*[width=9cm]{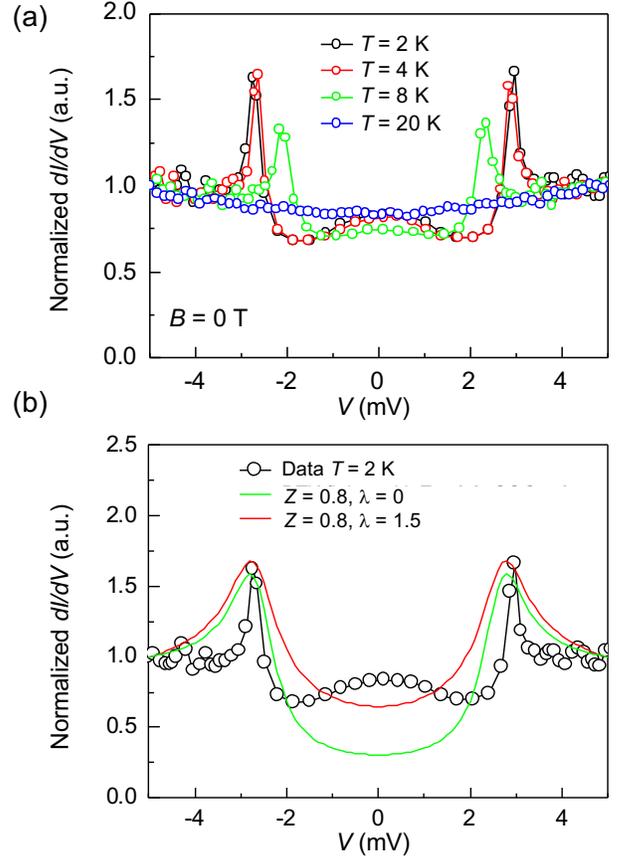}
\caption{(a) Measured and (b) calculated finite-temperature and $B=0$ differential conductance
for Fe$_{0.29}$TaS$_2$/NbN junction, normalized to its value at $V=5\,$mV, corresponding to the normal-state result. 
For $T=2\,$K comparison, in our calculations we choose $T/T_c=0.16$, $P=0.42$, and $\Delta=2.7\,$meV.
}
\label{fig:dIdV}
\end{figure}

We now turn to the measured bias-dependence for differential conductance, $dI/dV$, for a quasi-2D vdW F/S junction [recall its geometry in  Fig.~\ref{fig:FS}], 
shown in Fig.~\ref{fig:dIdV}(a) for several $T$. 
Some trends are expected, $dI/dV$ peaks reflect the peaks in the density of states,
near $|eV|=\Delta$~\cite{Blonder1982:PRB}, moving inward with $T$ as $\Delta(T)$ is also suppressed with $T$. 
While the perfect transmission in 1D N/S geometry near $V=0$ implies the doubling of the normal-state conductance~\cite{Blonder1982:PRB}, 
since an incident electron through Andreev reflection contributes to the transfer of an electron pair across the N/S interface, the F/S transmission 
should be diminished, both by the interfacial barrier, parametrized by $Z$ [recall Eq.~(\ref{eq:Z})] as well by the
spin polarization, $P=\Delta_\text{xc}/\mu_F \approx 0.4-0.5$, expected for Fe$_{0.29}$TaS$_2$
and measured in a similar Fe$_{0.26}$TaS$_2$ ferromagnet~\cite{Yu2019:APS}.  
Only a fraction of the incident majority spin electrons can
find the opposite spin partner for Andreev reflection~\cite{Soulen1998:S,Zutic2000:PRB}.

With the NbN critical temperature, $T_c=12.5\,$K, we focus on $dI/dV$ at the lowest $T=2\,$K and theoretically
explore the role of interfacial properties using dimensional parameters for barrier strength and Rashba SOC
\begin{eqnarray} 
Z &=&V_0 d \sqrt{m_F m_S}/(\hbar^2 \sqrt{k_F q_F}),   \\
\lambda &=& 2\alpha \sqrt{m_F m_S}/ \hbar ^2, 
\label{eq:dimless}
\end{eqnarray}
since we are interested in identifying trends with these interfacial parameters, rather than obtaining the best
fit for $dI(V)/dV$, we further simplify this approach by considering  the case for $m_F=m_S=m$ and $k_F=q_F$. 

Our calculated results are given 
in Fig.~\ref{fig:dIdV}(b) and reflect these simplifications. 
For example, assuming that, $k_F=q_F$ and that $m$ 
is the free electron mass, the chosen dimensionless barrier strenght $Z=0.8$ 
corresponds to  $k_F=1.4\times 10^{10}\,$m$^{-1}$,
consistent with NbN values~\cite{Chockalingam2008:PRB}, while 
$V_0=0.5\,$eV, and $d=1.7\,$nm is an average thickness among our studied F/S samples.
The value of Rashba SOC, $\alpha=5.7\,$eV\AA$^2$,
which follows from the choice of $\lambda=1.5$,
is also consistent with separate fits of the angular dependence of MR measured near zero bias in Ref.~\cite{Cai2021:NC}.

By comparing $\lambda=0$ and $\lambda=1.5$ results at a fixed $Z=0.8$ 
above the gap the changes are very small. However,  the inclusion of Rashba SOC 
can strongly enhance the low-bias conductance. To understand the origin of this SOC
effect, and similar trends from Sec.~II, we note that in the normal state 
for the barrier region the dispersion is SOC split and shifted 
by the barrier potential 
(assuming $V_0>0$).
As in Sec.~II, a spinor of an incident electron with $\bm{k}_\parallel$ can be decomposed into barrier
eigenspinors, $\left | \chi_\sigma \right\rangle  =  \left\langle \chi_+  {\left |\right.}\chi_\sigma \right\rangle 
\left | \chi _+ \right\rangle+\left\langle \chi _-  {\left |\right.} \chi _\sigma \right\rangle \left | \chi _- \right\rangle$,   
where $\chi_\pm$ from Eq.~(\ref{eq:Bspinor}) has  helicity $\pm1$. 
The two helicities have {\em inequivalent} effective barriers [recall Eqs.~(\ref{eq:Zout}) and (\ref{eq:Zin})] 
\begin{equation}
Z_\pm = Z \pm \lambda k_\parallel/2k_F.  
\label{eq:Zeff}
\end{equation}
Since $Z, \lambda k_\parallel/(2 k_F) \geq 0$, for positive helicity the barrier is enhanced, $Z^+_{\rm{eff}} \geq Z$.  
However, for negative helicity, at $Z = \lambda k_\parallel/(2 k_F)$, $Z_-$ becomes effectively completely transparent,
such $k_\|$ can be viewed as ``open channels" and are responsible for a strongly increased $dI/dV$.

Some peculiar conductance trends 
can already be understood at $V=0$ and $T=0\,$K,
where the charge transport in F/S junctions is determined by conventional and spin-flip Andreev reflection 
with opposite and equal spin projection of the incident and reflected particle, respectively, 
corresponding to the spin-singlet and spin-triplet interfacial superconducting correlations. 
For $G(V=0)$ plotted in the plane defined by $Z>0$ and $\lambda>0$, a striking behavior was found for $|P|<1$
in the triangular region approximately delimited with lines T1 and T2~\cite{Vezin2020:PRB}
\begin{equation}
\text{T1:}  \:  \: \lambda=2Z/\sqrt{1-P}, \quad \quad \text{T2:} \: \: \lambda=2Z,
\label{eq:L12}
\end{equation}
where $G$ is dominated by the spin-flip Andreev reflection and proximity-induced spin-triplet superconductivity.  

Before showing next that a region in $G$ delimited by the lines T1 and T2 also pertains to the enhanced calculated MAAR,
we 
comment on relaxing the assumption of equal masses and Fermi velocities in the F and S regions.
In a simple 1D N/S case, without SOC, different Fermi velocities, $F_v=(v_S/v_N)\neq1$, are known to increase the effective 
barrier strength, $Z_\text{eff}^2\rightarrow Z^2+(1-F_v)^2/4F_v > Z^2$~\cite{Blonder1983:PRB}. While this argument is often
used to also ignore a Fermi velocity mismatch in the F/S case~\cite{Soulen1998:S}, as it is accounted for
by simply enhancing $Z$, some subtleties exist~\cite{Zutic2000:PRB,Zutic1999:PRBa}. The mismatch of effective masses, $F_m=m_S/m_F$,
together with the mismatch of the Fermi wave vectors, $F_k= q_F/k_F$,
can then be viewed as determining $F_v$ and
enhancing the effective $Z$. In our studies with SOC, $F_k \neq 1$ implies that Eq.~(\ref{eq:Zeff}) should be generalized
by replacing $k_F$ with $\sqrt{k_F q_F}$,
including in the expression for open channels and a vanishing $Z_-$. 
Even with such a wave vector mismatch, the spin-triplet contribution remains enhanced within a triangular region, as long as $F_k > \sqrt{1-P}$ i.e., $q_F > k_\downarrow $. The delimiting lines T1 and T2, for both $F_k <1$ and $F_k>1$, have slopes modified by $F_k$ as compared to those given by Eq.~(\ref{eq:L12}).

\section{V. F/S Out-Of-Plane MAAR} 
Just as in the F/2DEG structures, the interplay between exchange field and Rashba SOC will lead to various nonmonotonic trends~\cite{Hogl2015:PRL,Vezin2020:PRB,Cai2021:NC,Mizuno2009:PRB}.
Such trends can be inferred from Eqs.~(\ref{eq:Zeff}) and (\ref{eq:L12}),
as well as in the MR, which in F/S junctions corresponds to the superconducting counterpart of TAMR,
and its term, MAAR (recall Sec.~I), identified the key role of the Andreev reflection.

Similar to the significance of the condition for perfect transmission and contour matching in understanding the origin of
an enhanced TAMR, for MAAR it is important to identify the influence of open channels~\cite{Vezin2020:PRB}.
From Eq.~(\ref{eq:Zeff}), we see that $Z, \lambda >0$ yields vanishing $Z_-$ with open channels
located at the Fermi contour of radius 
\begin{equation}
k_\|^\text{open}=(2Z/\lambda)k_F.
\label{eq:open}
\end{equation}
Open channels give the dominant contribution to $G$ and are also expected to determine the amplitude of MAAR.

\begin{figure}[t]
	\centering
	\includegraphics*[trim=0cm 0.1cm 0cm 0.1cm,clip,width=6.5cm]{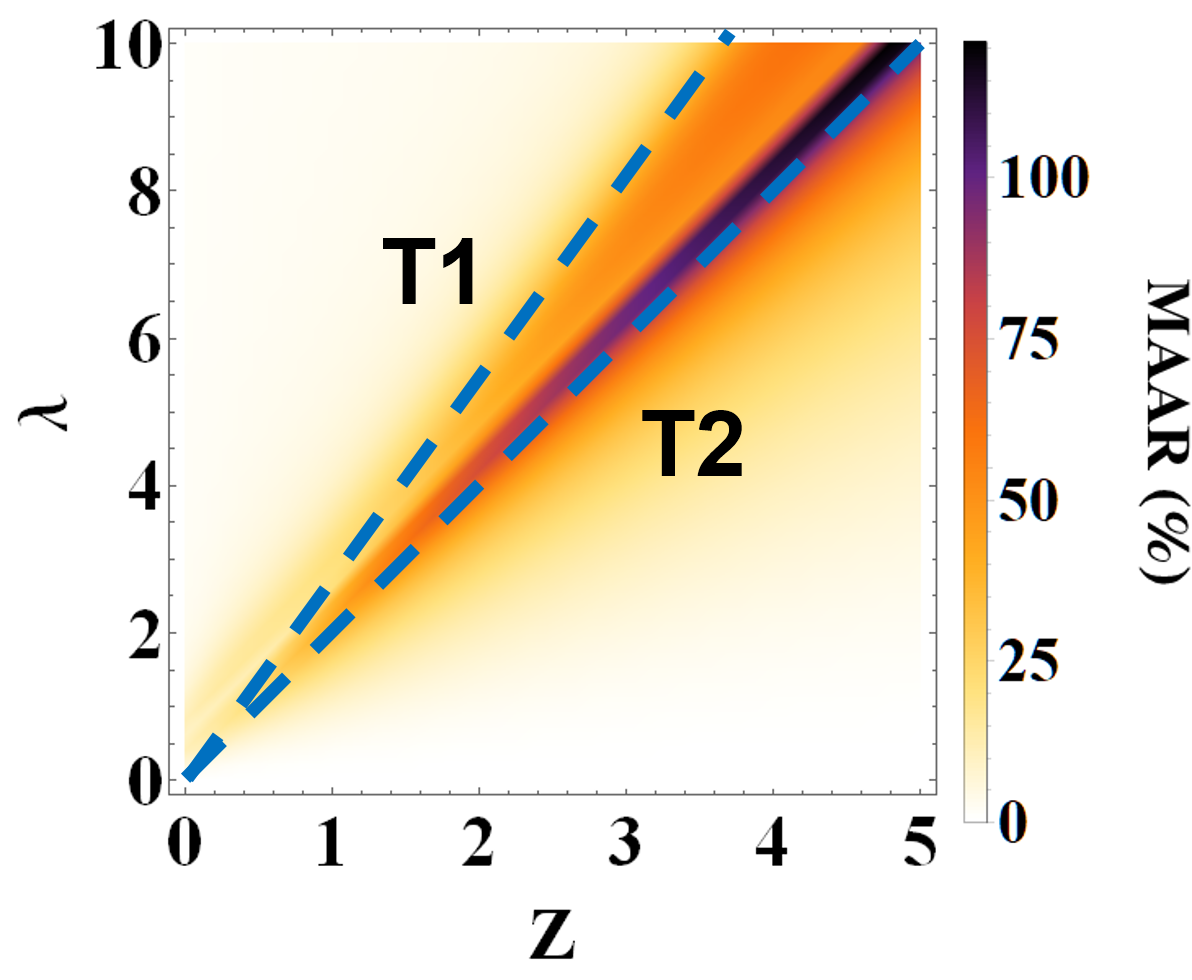}
	\caption{Amplitude of out-of-plane magnetoanisotropic Andreev reflection (MAAR), given by Eq.~(\ref{eq:tamrOUT})
		calculated at zero bias and $T=0\,$K, shown
		as a function of interfacial parameters $Z$ and $\lambda$ for
		$P = 0.47$, consistent with Fe$_{0.29}$TaS$_2$.
		Lines T1 and T2 are defined in Eq.~(\ref{eq:L12}).}
	\label{fig:MAAR}
\end{figure}

Since evanescent state for large $k_\|$ do not contribute to $G$ they should be excluded in open channels. 
From the Snell's law~\cite{Zutic2000:PRB}, for the incident majority spin electron this  implies that
 to have propagating scattering states, $k_\parallel \leq k_\downarrow$ is required for conventional Andreev reflections and 
$k_\parallel \leq q_F$ for dominant spin-flip Andreev reflections~\cite{Zutic1999:PRBa}.
The maximum of the total $G$ is achieved when the amount of the open channels 
$\propto k_\parallel$, is maximized. For $k_F = q_F$, in the two limiting case
$k_\|^\text{open}=k_\downarrow$ and $k_\|^\text{open}=k_F$ we recover the conditions for lines T1 and T2
from Eq.~(\ref{eq:L12}), such that the maximum spin-flip Andreev reflection is located near 
$k_\|^\text{open}=k_F$ i.e., $\lambda=2Z$.

From the calculated MAAR in Fig.~\ref{fig:MAAR} we see that the triangular region delimited by lines T1, T2
is also identifying the region of an enhanced MAAR and confirming the influence of open channels. Similar to our
contour matching in TAMR, as  noted in Sec.~II., for a full picture 
the role of spin mismatch also needs to be included. Two characteristic features are easy to see from Fig.~\ref{fig:MAAR}:
(i) a large MAAR enhancement, which can reach 100$\,$\% and (ii) MAAR is nonomontonic with both $Z$ and $\lambda$,
when the other parameter is kept constant.

To understand (i), we recall that in the triangular region,  the dominant contribution to the total $G$ is from the open channels 
that is associated with spin-flip Andreev reflection and spin-triplet pairing $\left| { \uparrow  \uparrow } \right\rangle$. 
Therefore, the total conductance can be approximately written as

\begin{eqnarray}
G (V=0) &\approx& \int {{k_\parallel }d{k_\parallel }\int_0^{2\pi } {\frac{{d\gamma }}{{\pi k_F^2}}R_{\sigma  = 1}^h\left( {{k_\parallel },\gamma } \right)} }  \\ \nonumber
&\approx& k_\parallel ^\text{open}\Delta {k_\parallel }\int_0^{2\pi } {\frac{{d\gamma }}{{\pi k_F^2}}R_{\sigma  = 1}^h\left( {k_\parallel ^\text{open},\gamma } \right)}
\label{eq:100}
\end{eqnarray}
where $\Delta {k_\parallel }$ is the width of the open channels in $\bm{k_\parallel}$-plane,  
$\gamma$ is the angle between $\bm{k_\|}$ and $\hat{\bm{k}}_x$, recall Eq.~(\ref{eq:Bspinor}).
For out-of-plane {\bf M}, $R_{\sigma  = 1}^h\left( {k_\parallel ^\text{open},\gamma } \right) \equiv R_{\sigma  = 1}^{h,\text{open}}$ does not depend on $\gamma$ 
due to the rotational symmetry and we can write
\begin{equation}
G\left( {\theta  = 0} \right) \approx \frac{2k_\parallel^\text{open}\Delta {k_\parallel}}{k_F^2} R_{\sigma  = 1}^{h,\text{open}}.
\label{eq:Gout}
\end{equation}

\begin{figure}[t]
\centering
\includegraphics*[trim=0.6cm 2.4cm 0.cm 2.4cm,clip,width=9cm]{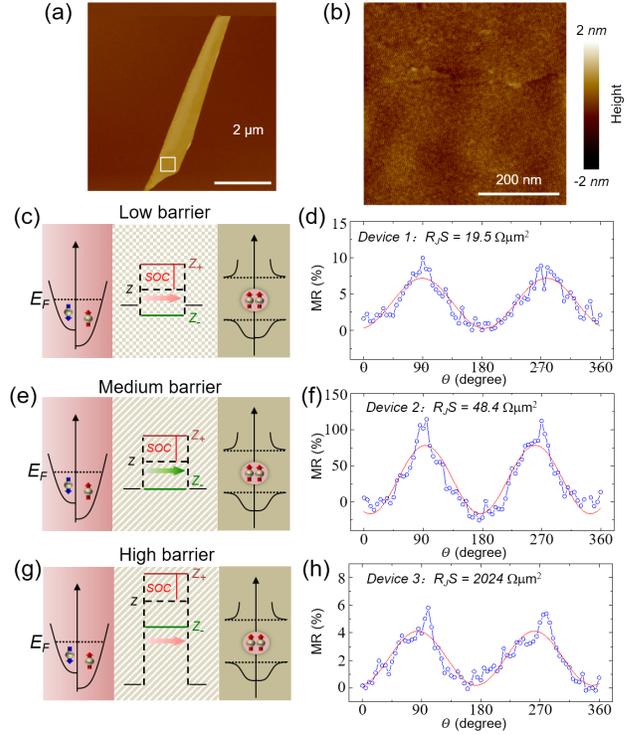}
\caption{(a) and (b) Atomic force microscope images of 2$\,$nm Al$_2$O$_3$ on Fe$_{0.29}$TaS$_2$ thin flake with RMS roughness
of $\sim0.270\,$nm indicate that Al$_2$O$_3$ is relatively flat and without obvious pin holes or discontinuities. 
(c), (e), and (g) schematic of the incident spin-polarized electrons tunneling across different effective interfacial 
barriers with Rashba SOC into the proximity-induced spin-triplet superconductivity. (d), (f), and (h) the corresponding 
magnetoresistance (MR) [recall $\theta$ in Fig.~\ref{fig:FS}], for different  junction resistance-area products $R_J S$,
a proxy for dimensionless barrier strength $Z$. Red lines: theoretically predicted two-fold symmetry.
The measurements were at $T=2\,$K, $B=9\,$T, and $V=1\,$mV.
}
\label{fig:MR}
\end{figure}

For in-plane  {\bf M}, considering the interplay between the incident electron spin and the barrier eigenspinor, 
the spin-flip Andreev reflection of  an incident $\uparrow$ electron is given by 
$R_{\sigma  = 1}^h\left( {k_\parallel^\text{open},\gamma } \right) \approx R_{\sigma  = 1}^{h,\text{open}}{\cos ^2}\left( {\gamma  - \phi } \right)$~\cite{Vezin2020:PRB}. 
As a result, for in-plane  {\bf M}, we obtain
\begin{equation}
G\left( \theta  = \frac{\pi}{2} \right) \approx \frac{k_\parallel^\text{open}\Delta {k_\parallel}}{k_F^2}  R_{\sigma  = 1}^{h, \text{open}} = \frac{G\left( {\theta  = 0} \right)}{2}.
\label{eq:Gin}
\end{equation}
From Eq.~(\ref{eq:tamrOUT}), this means that MAAR in the triangular region is 100\%, which explains that the previous prediction~\cite{Hogl2015:PRL} 
can be attributed to spin-triplet superconductivity and is also consistent with the largest values
in the experimental results  for Fe$_{0.29}$TaS$_2$/NbN junctions
with a perpendicular magnetic anisotropy~\cite{Cai2021:NC}.

Despite our simple theoretical model to include the interplay of exchange field and SOC on the proximity-induced superconductivity in F/S junctions,
such a description provides MAAR which is in a good agreement with the measured angular dependence of MR in Fig.~\ref{fig:MR}.
Remarkably, just as the variation of $Z$ shows a nonmonotonic MAAR in Fig.~\ref{fig:MAAR} and a huge increase of MAAR at 
intermediate $Z$ values, which with Rashba SOC can lead to a vanishing $Z_-$, we see that this trend in Figs.~\ref{fig:MR}(d), (e), and (h) is also retained 
in the measured results obtained by varying the Al$_2$O$_3$ barrier thickness ($\sim1-2.5\,$nm).

While we cannot rule out a non-MAAR contribution in the measured low-bias MR for Fe$_{0.29}$TaS$_2$/NbN junctions, we can still examine its magnitude
from several reported control measurements~\cite{Cai2021:NC}. For example,  by increasing $T$ above $T_c$, MR is drastically reduced from its 
maximum at $\sim 100\,$\% in Fig.~\ref{fig:MR}(f)
and we can attribute the main MR part to a superconducting response. By considering MR in Al/Al$_2$O$_3$/NbN samples, we no longer have
the interplay between the ferromagnetism and SOC, responsible for MAAR, but the resulting small MR of just few $\%$ at $T=2\,$K is encouraging
that the presence of vortices at $B=9\,$T has only a modest MR effect. 
Taken together, these findings confirm that MAAR is indeed the dominant source of the measured MR in Fig.~\ref{fig:MR}, which further indicates 
that F$_{0.29}$TaS$_2$/NbN junctions, provide an important platform to realize elusive equal-spin-triplet superconductivity. 

\section{VI. Discussion and Conclusions}

We have shown that Rashba SOC can lead to strongly enhanced magnetoresistance in junctions with one ferromagnet.
Both in the normal-state and the superconducting response the calculated magnetoresistance is characterized 
by various nonmonotonic trends. 
While some of these trends and an enhanced magnetoresistance have been measured in
superconducting junctions~\cite{Martinez2020:PRA,Cai2021:NC}, experiments in the normal state are largely unexplored.

To realize magnetic proximity effects for the in-plane transport, magnetic insulators are desirable~\cite{Zutic2019:MT,Jiang2015:NL,Wei2013:PRL,Lee2014:PRB}. 
This precludes current flow in the more resistive F region [Fig.~1(a)] and minimizes hybridization with the 2DEG to
enable a gate-tunable proximity-induced exchange splitting in their respective states. However, as shown in graphene~\cite{Lazic2016:PRB,Zollner2016:PRB,Xu2018:NC} 
for tunable magnetic proximity effects one could also employ ferromagnetic metals, separated by an insulating layer from the 2DEG .

For a suitable materials platform, which would support large magnetoresistive effects, we
could extend our focus on simple Rashba SOC to a growing class of van der Waals (vdW) materials. 
Their heterostructure offer both transport in materials with strong SOC as well 2D ferromagnets~\cite{Gong2017:N,Huang2017:N,Gong2019:S}
with atomically-sharp interfaces and not limited to lattice-matching constraints.
For example, transition-metal dichalcogenides in addition to their inherent SOC also provide spin-orbit 
proximity~\cite{Avsar2014:NC,Yan2016:NC, Han2014:NN,Dankert2017:NC,Wang2015:NC,AntonioBenitez2018:APLM,Douli2020:NL}
and thereby alter spin textures and expected TAMR, while 2D ferromagnets support 
a versatile gate control~\cite{Burch2018:N,Deng2018:N,Xing2017:2DM}. 

While we have focused on a longitudinal transport in a very simple system, the 
behavior emerging from a spin parity-time symmetry of the scattering states with perfect
transmission
is important not just in explaining a surprisingly large TAMR, 
but also as a sensitive probe to distinguish between trivial and topological states~\cite{Shen2020:PRB}.
It would be interesting to know if our predicted nonmonotonic trends with interfacial parameters for TAMR could
be also relevant for other transport phenomena in junctions with a single ferromagnet, such as spin-orbit torque and spin-Hall magnetoresistance~\cite{Tsymbal:2019,Belashchenko2020:PRB}.  

In the superconducting state, the observed large magnetoresistance has important implications as it provides
an alternative probe for spin-triplet or mixed singlet-triplet superconductivity~\cite{Gorkov2001:PRL} and such a large signal is  possible to realize even 
systems with only a modest SOC and negligible normal-state magnetoresistance~\cite{Martinez2020:PRA}.
Since this work shows that spin-triplet superconductivity, desirable both for superconducting spintronics
and Majorana bound states~\cite{Amudsen2022:P}, is feasible in simple structures with a single F and S region, 
it would also be important to extend its analysis to Josephson junctions, where enhanced magnetoresistance was
predicted~\cite{Costa2017:PRB}, but not connected to the spin-triplet superconductivity, which was extensively studied
in other Josephson junctions~\cite{Khaire2010:PRL,Robinson2010:S,Banerjee2014:NC,Valls:2022,Cai2022:AQT,Amudsen2022:P}.

In our work, the nonmonotonic trends with the interfacial barrier strength were observed
by comparing samples with different thickness of the (Al$_2$O$_3$). It would be desirable, to realize systems
in which such changes, as well as the tunability of the Rashba SOC strength could be controlled in the same sample. 
A desirable progress is realized in Josephson junctions where currently there are no ferromagnetic regions, 
but the Zeeman splitting is due to an applied magnetic field~\cite{Amudsen2022:P,Fornieri2019:N,Ren2019:N,Dartiailh2021:PRL,Zhou2022:NC,Banerjee2022:X}.
Related experimental support in junctions with Al as a superconductor and InAs-based 2DEG already suggests an 
observation of topological spin-triplet superconductivity~\cite{Dartiailh2021:PRL}.
In the same sample that both a reentrant superconductivity with an applied magnetic field and the jump in the
superconducting phase are measured, but an enhanced magnetoresistance as another signature
of spin-triplet superconductivity has not yet been studied.

\section{Acknowledgments}
We thank Thomas Vezin for useful discussions. We had a privilege of first meeting Professor Emmanuel Rashba at the Spintronics 2001:
International Conference on Novel Aspects of Spin-Polarized Transport and Spin Dynamics in Washington, D.C. (chaired
by one of us~\cite{Zutic2002:JS}). He gave an insightful talk sharing his personal experience about relevant historic developments as well as
a vision for the future 
opportunities in spintronics, conveyed further in his article~\cite{Rashba2002:JS}.
Since then, through his valuable comments, we were able to better understand our own work
and its broader context. His advice and guidance was not limited to semiconductor systems, and we again benefited 
from his comments which came soon after posting online our recent manuscript on spin-triplet superconductivity~\cite{Cai2021:NC}.  
This work is supported by U.S. Department of Energy, Office of Science, Basic Energy Sciences under Award No. DE-SC0004890,
the UB Center for Computational Research, National Basic Research Programs of China (No. 2019YFA0308401), 
and National Natural Science Foundation of China (No. 11974025).

\bibliography{MR_Rashba}
\end{document}